\documentclass[twocolumn,aps,showpacs,preprintnumbers,byrevtex,superscriptaddress]{revtex4}
\usepackage{graphicx}
\usepackage{dcolumn}
\usepackage{bm}

\begin{document}

\preprint{Squeezing in NCQSE}
\title{Quantum Fluctuations around Bistable Solitons in the Cubic-Quintic
nonlinear Schr\"{o}dinger equation}
\author{Ray-Kuang Lee}
\affiliation{Institute of Electro-Optical Engineering, National Chiao-Tung University, Hsinchu, Taiwan}
\author{Yinchieh Lai}
\email{yclai@mail.nctu.edu.tw}
\affiliation{Institute of Electro-Optical Engineering, National Chiao-Tung University, Hsinchu, Taiwan}
\author{Boris A. Malomed}
\email{malomed@eng.tau.ac.il}
\affiliation{Department of Interdisciplinary Studies, School of Electrical Engineering, Faculty of Engineering, Tel Aviv University, Tel Aviv 69978, Israel}
\date{\today }

\begin{abstract}
Small quantum fluctuations in solitons described by the cubic-quintic nonlinear Schr\"{o}dinger equation (CQNLSE) are
studied with the linear approximation.
The cases of both self-defocusing and self-focusing quintic term are considered (in the latter case, solitons may be effectively stable, despite the possibility of collapse).
The numerically implemented back-propagation method is used to calculate the optimal squeezing ratio for the quantum fluctuations vs. the propagation distance.
In the case of the self-defocusing quintic nonlinearity, opposite signs in front of the cubic and quintic terms make the fluctuations around bistable pairs of solitons (which have different energies for the same width) totally different.
The fluctuations around nonstationary Gaussian pulses in the CQNLSE model are studied too.
\end{abstract}

\pacs{42.50.Lc, 42.65.Tg, 42.81.Dp, 42.65.Pc}
\keywords{Quantum fluctuations, quantum noise, and quantum jumps, Optical
solitons; nonlinear guided waves, Propagation, scattering, and losses;
solitons, Optical bistability, multistability, and switching, including
local field effects}
\maketitle


\section{Introduction}
It is well known that generalized nonlinear Schr{\"{o}}dinger equation (NLSE) with a saturable response can support multi-stable temporal or spatial solitons in one- and multi-dimensional cases \cite {Edmundson, Enns,saturable-bistable,Kaplan85a,Kaplan85b,Herrmann92}.
The saturation of the Kerr nonlinearity may also prevent collapse of optical spatio-temporal solitons (\textquotedblleft light buttes\textquotedblright ), in two- and three-dimensional cases \cite{Desyatnikov00, Mihalache03}.
The general form of the NLSE with saturable nonlinearity is (in the temporal domain) 
\begin{equation}
iU_{z}+U_{tt}+\mathcal{F}(|U|^{2})U=0
\label{NLS}
\end{equation}
where $U(z,t)$ is the local amplitude of the electromagnetic wave, $z$ and $t $ are, as usual, the propagation distance and reduced time, and the function $\mathcal{F}(|U|^{2})$ describes the saturable nonlinear response of the medium.
The simplest form of the latter function corresponds to the cubic-quintic (CQ) nonlinearity, with the cubic and quintic terms accounting for the self-focusing and self-defocusing, $F(|U|^{2})=|U|^{2}-b|U|^{4}$, where $b$ is the ratio of cubic and quintic nonlinearity.
Despite the obvious danger of the spatio-temporal collapse, we will demonstrate that reasonable results may be obtained in the latter case too.

The corresponding cubic-quintic NLSE (to be abbreviated CQNLSE) describes light propagation in chalcogenide glasses \cite{glass} and organic media \cite{organic}, if nonlinear absorption may be neglected.
In fact, CQNLSE is the simplest model that makes it possible to study quantum fluctuations around bistable solitons (and non-soliton Gaussian pulses) in systems with saturable nonlinearity, therefore results reported below may apply to a broader class of optical media than those which are directly modeled by the combination of cubic and quintic terms (although stability of solitons is not the same in models with different forms of the saturable nonlinearity \cite{Kolokolov}).

The theory of quantum fluctuations around optical solitons has been developed during the past 15 years by means of several techniques, which were applied to the family of solitons in the NLSE model \cite{Drummond87, Lai89a, Lai89b}, higher-order solitons in the same system \cite{Schmidt00}, self-induced-transparency solitons \cite{Lai90}, and Bragg solitons \cite{Lee04}.
However, quantum theory has not yet been worked out for bistable solitons, which we aim to carry out in this work, using the CQNLSE model in the simplest (1+1)-dimensional case, and applying the known \textit{back-propagation method} \cite{YLai95}.
By means of this method, we investigate quantum fluctuations as a function of the distance passed by the pulse.
A measurement scheme which should enable the observation of the predicted effects is the homodyne detection, which measures the inner product between the given pulse and a local-oscillator gauge pulse, according to the projection interpretation of the homodyne detection \cite{Haus90}.

First of all, we find that the optimal squeezing ratio of the quantum fluctuations around the pulse improves when both the cubic and quintic terms are of the focusing type; on the other hand, the squeezing ratio degrades when the cubic term is focusing and the quintic one is defocusing.
The quantum fluctuations around two solitons with equal widths, belonging to a bistable pair, are found to be totally different due to the effects of the quintic nonlinearity.
Quantum fluctuations around bistable Gaussian pulses (nonstationary ones) will also be studied to compare the results with those for the solitons.

This paper is organized as follows: in section \ref{secFML}, we derive the quantum CQNLSE model and the corresponding adjoint linearized equation for quantum fluctuations.
The effect of quintic nonlinearity on quantum fluctuations, and the differences between equal-width solitons belonging to a bistable pair, are highlighted in section \ref{secBST}.
Section \ref{secgauss} deals with the fluctuations around nonstationary Gaussian pulses, and comparison with the case of the solitons.
The paper is concluded by section \ref{secCCL}.

\begin{figure}[tbp]
\begin{center}
\includegraphics[width=3.0in]{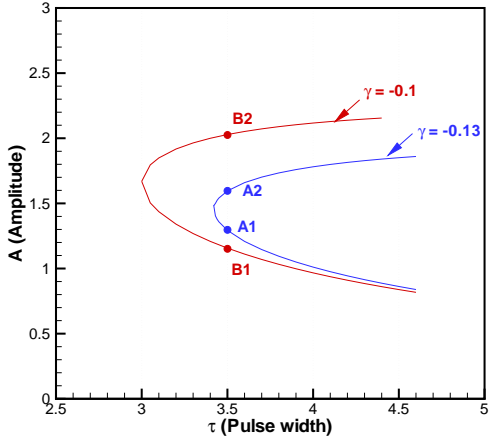}
\end{center}
\caption{The amplitude-pulsewidth relation for the bistable solitons as given by Eq. (\protect\ref{eqBST}). Two pairs of bistable solitons with equal pulsewidth, $A_{1,2}$ and $B_{1,2}$, are selected as examples for detailed investigation.}
\label{fig-map}
\end{figure}

\section{Quantum cubic-quintic nonlinear Schr{\"{o}}dinger equation}
\label{secFML}
The cubic-quintic version of the NLSE (\ref{NLS}) is 
\begin{equation}
iU_{z}+U_{tt}+2\chi |U|^{2}U+3\gamma |U|^{4}U=0,  \label{NCQSE}
\end{equation}
where the cubic coefficient is normalized to be $\chi =\pm 1$.
In the case of $\chi >0$, exact solutions for bistable solitons are available \cite {Kaplan85a, Herrmann92},
\begin{equation}
U(z,t)=\sqrt{\frac{2\beta }{\sqrt{1+4\gamma \beta }\cosh \left( 2\sqrt{\beta}t\right) +\chi }}e^{i\beta z}  \label{sech}
\end{equation}
where $\beta $ is a propagation constant (intrinsic parameter of the soliton family), subject to the condition $1+4\gamma \beta >0$.
The peak power of the solution (\ref{sech}), $A^{2}=2\beta /\left( \sqrt{1+4\gamma \beta }+\chi \right) $, is related to the full-width at half-maximum pulse width , $\tau $, by the following relation:
\begin{equation}
\cosh \left(0.5\tau \sqrt{\chi A^{2}+\gamma A^{4}}\right) =\frac{3\chi+4\gamma A^{2}}{\chi +2\gamma A^{2}}.  \label{eqBST}
\end{equation}

The amplitude-pulsewidth relation for the bistable solitons, corresponding to Eq. (\ref{eqBST}), is displayed in Fig. \ref{fig-map} for $\chi =+1$ and $\gamma <0$.
As is seen, the pulsewidth of the soliton cannot be smaller than a minimum (critical) value $\approx 3.0$ for $\gamma = -0.1$ and $\approx 3.42$ for $\gamma = -0.13$.
Two pairs of bistable solitons with identical pulsewidths, marked as $A$ (which is close to the turning point) and $B$ (taken farther from the turning point), are chosen to be the illustrative examples.
In particular, the pulsewidth of the bistable solitons belonging to Pair $B$ is $3.5$, and their amplitudes are $1.147$ and $2.033$, respectively.
Solitons belonging to both branches of the curve in Fig. \ref{fig-map} relation are stable as classical solutions.

In the quantum theory, the classical CQNLSE (\ref{NCQSE}) is replaced by its quantum counterpart, with $U(z,t)$ directly replaced by operator field variable $\hat{U}(z,t)$:
\begin{equation}
\hat{U}_{z}=i\hat{U}_{tt}+2i\chi \hat{U}^\dag \hat{U}\hat{U}+3i\gamma \hat{U}^\dag \hat{U}^\dag \hat{U} \hat{U} \hat{U}.
\label{QNCQSE}
\end{equation}
The quantized field must satisfy the equal-coordinate Bosonic commutation relations,
\begin{eqnarray}
&&\left[ \hat{U}(z,t_{1}),\hat{U}^{\dag }(z,t_{2})\right] =\delta(t_{1}-t_{2}),  \nonumber \\
&&\left[ \hat{U}(z,t_{1}),\hat{U}(z,t_{2})\right] =\left[ \hat{U}^{\dag}(z,t_{1}),\hat{U}^{\dag }(z,t_{2})\right] =0  \label{comm}
\end{eqnarray}
Equation (\ref{QNCQSE}), which provides for the Heisenberg-picture description, can be derived from the Hamiltonian
\[
\hat{H}=-\int d\,z\left( \hat{U}^{\dag }\frac{\partial ^{2}}{\partial t^{2}}
\hat{U}+\chi \hat{U}^{\dag }\hat{U}^{\dag }\hat{U}\hat{U}+\gamma \hat{U}
^{\dag }\hat{U}^{\dag }\hat{U}^{\dag }\hat{U}\hat{U}\hat{U}\right) ,
\]
so that Eq. (\ref{QNCQSE}) is tantamount to $i\hat{U}_{z}=[\hat{U},\hat{H}]$ (the replacement of the ordinary equal-time correlation relations for quantum fields by the equal-coordinate ones can be justified, in the present physical context, within the framework of the canonical approach, as shown in Ref. \cite{Matsko}).

Next, we substitute $\hat{U}=U_{0}+\hat{u}$ into Eq. (\ref{QNCQSE}), to linearize the equation around the classical solution $U_{0}$ for the soliton containing a very large number of photons.
The linearized equation for the quantum fluctuations reads
\begin{eqnarray}
\hat{u}_{z} &=&i\hat{u}_{tt}+4i\chi |U_{0}|^{2}\hat{u}+9i\gamma |U_{0}|^{4}\hat{u}  \nonumber \\
&+&2i\chi U_{0}^{2}\hat{u}^{\dag }+6i\gamma U_{0}^{3}U_{0}^{\ast }\hat{u}^{\dag }\nonumber \\
&\equiv& \hat{\mathcal{P}}~\hat{u},
\label{linear1}
\end{eqnarray}
where $\hat{\mathcal{P}}$ is an effective evolution operator.
The operator perturbation field $\hat{u}$ satisfies the same equal-coordinate Bosonic commutation relation (\ref{comm}) as the unperturbed field operator $\hat{U}$.

To describe the quantum fluctuations, we need to find the corresponding adjoint field which satisfies the condition
\begin{equation}
\langle u^{A}|\hat{\mathcal{P}}\hat{u}\rangle =\langle \mathcal{P}^{A}u^{A}| \hat{u}\rangle ,  \label{eqrqr}
\end{equation}
where the inner product is defined by
\begin{equation}
\langle f|\hat{g}\rangle =\frac{1}{2}\int_{-\infty }^{+\infty }\left(f^{\ast }\hat{g}+f\hat{g}^{\dag }\right) dt.
\label{product}
\end{equation}
This definition of the inner product conforms to the principle that any physical observable can be expressed as the inner product between a characteristic measurement function and the quantum-field operator of the perturbation \cite{YLai95}.
The adjoint field which implements the condition (\ref{eqrqr}) obeys the following linear operator equation:
\begin{eqnarray}
u_{z}^{A} &=&iu_{tt}^{A}+4i\chi |U_{0}|^{2}u^{A}+9i\gamma |U_{0}|^{4}u^{A}\nonumber \\
&-&2i\chi U_{0}^{2}u^{A\ast }-6i\gamma U_{0}^{3}U_{0}^{\ast }u^{A\ast}\nonumber \\
&\equiv& -\mathcal{P}^{A}~u^{A}.
\label{linear2}
\end{eqnarray}
It can be checked that the inner product (\ref{product}) between solutions of Eqs. (\ref{linear1}) and (\ref{linear2}) is preserved in the evolution.
Using this invariance, one can express the inner product, between the quantum perturbation field and a properly chosen projection function, at an output point, $z=L$, in terms of the input quantum perturbation at the initial point $z=0$,
\[
\langle u^{A}(z=L,t)|\hat{u}(L,t)\rangle =\langle u^{A}(0,t)|\hat{u}
(0,t)\rangle ,
\]
which is the basis of the back-propagation method.

Then, it is easy to calculate the quantum uncertainty of the output field, knowing the statistics of the input quantum-field operators.
In particular, the squeezing ratio of an observable $f$ is calculated as
\begin{equation}
R(L)\equiv \frac{\mathrm{var}[\langle f_{L}(t)|\hat{u}(L,t)]}{\mathrm{var} [\langle f_{L}(t)|\hat{u}(0,t)]}=\frac{\mathrm{var}[\langle f_{0}(t)|\hat{u} (0,t)]}{\mathrm{var}[\langle f_{L}(t)|\hat{u}(0,t)]}, \nonumber
\end{equation}
where $\mathrm{var}[\cdot ]$ means the variance, $f_{L}(t)$ is the projection function at the output point, and $f_{0}(t)$ is the back-propagated projection function.
In the case of the homodyne detection, the output is the inner product of the input field operator with the local oscillator state \cite{Haus90, Lai93}.
In the usual squeezing experiment, the measurement function $f_{L}(t)$ is a local oscillator pulse with following expression [recall $U_{0}(z,t)$ is the classical solution for the field pulse (in particular, soliton)],
\begin{equation}
f_{L}(t)=\frac{U_{0}(L,t)e^{i\theta }}{\sqrt{\int_{-\infty }^{+\infty}d\,t|U_{0}(L,t)|^{2}}}~,
\label{oscillator}
\end{equation}
where $\theta $ is an adjustable phase shift between the local oscillator and the signal pulse of the homodyne detection.
The optimal (minimum) vale of the squeezing ratio $R(T)$ can be chosen, varying the parameter $\theta $.

Finally, we assume the input quantum perturbation-field operator corresponds to a coherent state, which is good assumption in most cases.
Based on the formulation given above, in the next section we calculate the optimal squeezing ratios for solitons in the quantum CQNLSE model.

\section{Quantum fluctuations about solitons}
\label{secBST}

\subsection{Effects of quintic nonlinearity on quantum fluctuations}
\begin{figure}[tbp]
\begin{center}
\includegraphics[width=3.0in]{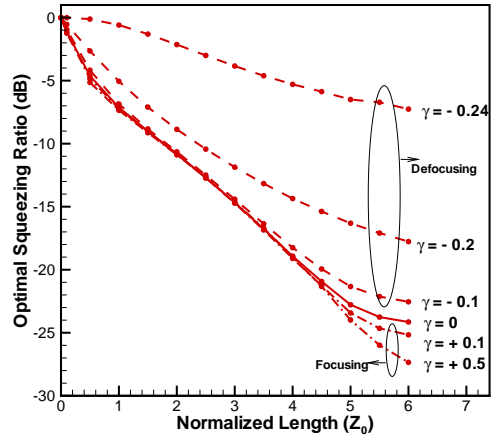}
\end{center}
\caption{The optimal squeezing ratio as a function of propagation distance for different values of the strength $\protect\gamma$ of the quintic nonlinearity, while the cubic-nonlinearity coefficient is fixed, $\chi = 1.0$.
When $\protect\gamma > 0$, squeezing is stronger, while for $\protect\gamma < 0$, it is weaker.}
\label{fig-gamma}
\end{figure}

To start the analysis of quantum fluctuations around the soliton, we fix the self-focusing cubic nonlinearity coefficient to be $\chi =+1$ and vary the quintic nonlinearity coefficient $\gamma $, in order to study effects of the quintic nonlinearity on quantum fluctuations.
The initial pulse is taken in the form of the soliton solution in equation (\ref{sech}).
The optimal squeezing ratio vs. the propagation distance is shown in Fig. \ref{fig-gamma}.
For $\gamma = 0 $, the CQNLSE is reduced to the usual cubic NLSE, for which quantum fluctuations have been studied in detail \cite{Lai90}.
If $\gamma >0$, i.e., both the cubic and quintic nonlinearities are focusing, the optimal squeezing ratio is almost the same as the case of $\gamma =0$, provided that the propagation distance is short enough.
That is because the quintic nonlinearity is not strong enough to affect CQNLSE soliton in a such short distance.
However, for longer distances ($_{\sim }^{>}~5\ $soliton periods), the effect of the quintic nonlinearity accumulates, improving the optimal squeezing ratio.
On the other hand, the quintic term with $\gamma <0$ is defocusing, causing degradation of the optimal squeezing ratio, due to partial compensation between the focusing (cubic) and defocusing (quintic) nonlinearities; if the defocusing effect is strong ($\gamma =0.2$), the optimal squeezing ratio quickly decreases even after propagating a short distance.

\begin{figure}[tbp]
\begin{center}
\includegraphics[width=3.0in]{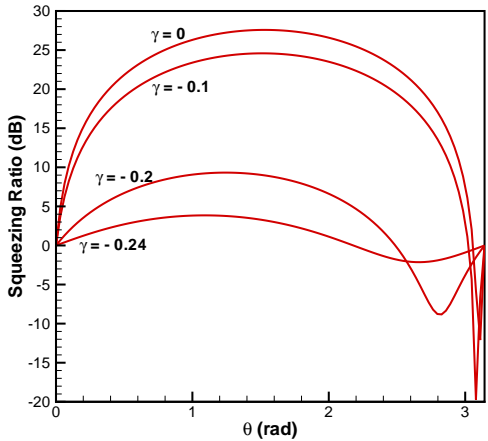}
\end{center}
\caption{The squeezing ratio as a function of the phase of the local oscillator $\protect\theta$ used for the homodyne detection.
In this case, $\protect\chi = +1.0$, $\protect\gamma < 0$.}
\label{theta}
\end{figure}

Figure \ref{theta} shows the dependence of the squeezing ratio on the phase of the local oscillator in the homodyne detection when the quintic nonlinearity is defocusing $\gamma < 0$.
This dependence indicates quadrature squeezing of the quantum fluctuations around the CQNLSE solitons, like the case of the ordinary NLSE solitons ($\gamma =0$).
With the self-focusing quintic term, $\gamma >0$, the CQNLSE solitons also undergo quadrature squeezing.

\subsection{Bistable solitons}
\begin{figure}[tbp]
\begin{center}
\includegraphics[width=3.0in]{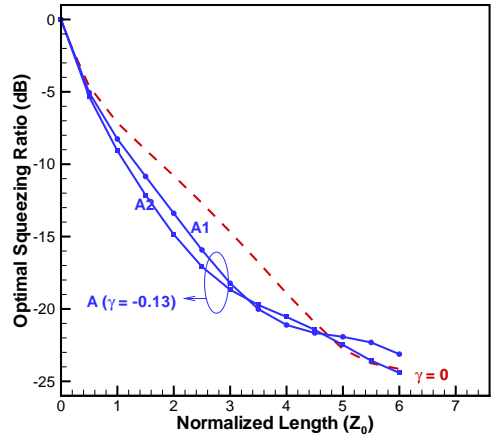}
\includegraphics[width=3.0in]{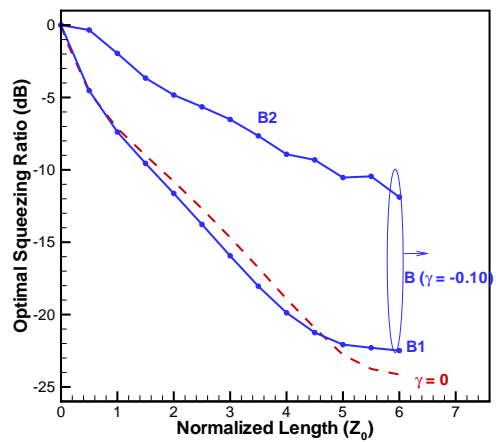}
\end{center}
\caption{The optimal squeezing ratios for two pairs of bistable solitons.
The case of the cubic nonlinear Schr\"{o}dinger equation, $\protect\gamma =0$, with the amplitude $A = 1.0$ is also plotted (dashed line) for comparison.}
\label{fig-bi-soliton}
\end{figure}

Here, we aim to compare the evolution of quantum fluctuations for bistable solitons belonging to the two pairs marked in Fig. \ref{fig-map}.
Recall that the pair $A$ is taken at $\gamma =-0.13$, which is close to its turning point, $\gamma =\gamma _{\mathrm{cr}}$, and pair $B$ is taken farther from its turning point, at $\gamma =-0.1$; the amplitudes of the solitons are, respectively, $A_{1}=1.3045$, $A_{2}=1.5531$, and $B_{1}=1.147$, $B_{2}=2.033 $.
The optimal squeezing ratios for these two pairs of the bistable solitons are shown in Fig. \ref{fig-bi-soliton}.

We observe that, for the solitons belonging to the pair $A$, which have the same widths, $\tau = 3.5$, and slightly different amplitudes, the evolution of the quantum fluctuation is similar.
One may expect that the soliton with a higher amplitude will be more squeezed, this is why the optimal squeezing ratio for the $A_{2}$ soliton is smaller than for the $A_{1}$one, in the beginning.
But due to the defocusing effect of the quintic nonlinearity (recall we now consider the case of $\gamma < 0$), the optimal squeezing ratios for both solitons degrade after propagating a certain distance (about $3$ and 4.5 soliton periods for $A_{2}$ and $A_{1}$, respectively), which leads to crossing of the squeezing-ratio curves for these bistable solitons.

For the pair $B$, the difference of the amplitudes in the soliton pair is large.
In accordance with this, we find that properties of the quantum fluctuations are totally different for these two solitons.
The curve of the squeezing ratio for the soliton with the smaller amplitude, $B_{1}=1.147$, is like the ordinary NLSE solitons (the dashed line in Fig. \ref {fig-bi-soliton}), while the squeezing for the soliton with the larger amplitude, $B_{2}=2.033$, is much poorer, which is naturally explained by the strong defocusing effect exerted on the latter soliton by the quintic self-defocusing term.

\section{Quantum fluctuations about nonstationary pulses}
\label{secgauss}
For situations where exact soliton solutions for pulses are not available, the variational approximation (VA) is known to be an efficient analytical method (see a review \cite{Malomed02}).
In particular, the VA for the classical CQNLSE solitons was developed in Ref. \cite{Angelis94}, using the Gaussian \textit{ansatz} for the pulse waveform,
\[
U(z,t)=A~\exp \left( -\left( t^{2}/2\alpha ^{2}\right) +iat^{2}\right) ,
\]
where $\alpha $ and $a$ are, respectively, the width and chirp of the pulse.
The use of the Gaussian makes sense not only because it is convenient for the application of VA, but also due to the fact that laser sources usually produce pulses with this shape (including the intrinsic chirp).

\begin{figure}[tbp]
\begin{center}
\includegraphics[width=3.0in]{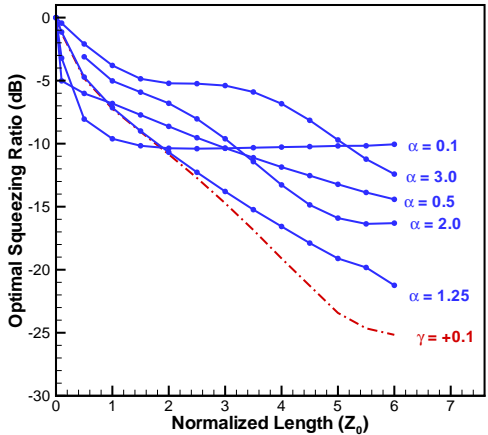}
\end{center}
\caption{The optimal squeezing ratio vs. the propagation distance for different nonstationary Gaussian pulses with different widths.
The result for the NCQSE soliton is plotted by the dashed-dotted line, for comparison.
In this figure, $\protect\gamma =+0.1$, and $A = 1.0$.}
\label{fig-alpha}
\end{figure}

Strictly speaking, the Gaussian ansatz may only produce a nonstationary pulse (plus some radiation).
Nevertheless, it is possible to calculate quantum fluctuation around it, and compare the results with those presented above for the solitons.
In general, the squeezing ratio for nonstationary pulses cannot be larger than for the soliton of the same width, due to emission of radiation waves by the nonstationary pulse.
Results of the calculation of the optimal squeezing ratio for the Gaussian pulses with different widths are displayed in Fig. \ref{fig-alpha}.
The Gaussian pulse produces a similar but poorer squeezing curve than the soliton of the same width, $\alpha =1.25$ (the dashed-dotted line, marked by \textquotedblleft  $\gamma =0.1$\textquotedblright\ in Fig. \ref{fig-alpha}).
Better squeezing ratios can be obtained on a short propagation distance for shorter pulses, with $\alpha <1.25$, but they degrade very quickly with the increase of the distance.
For broader pulses, with $\alpha >1.25$, the radiation modes will strongly affect the optimal squeezing ratio from the very beginning.

\begin{figure}[tbp]
\begin{center}
\includegraphics[width = 3.0in]{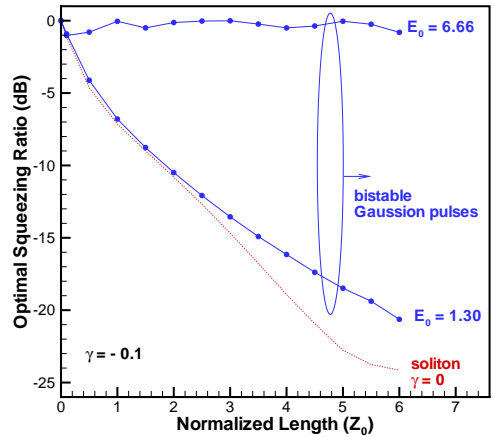}
\end{center}
\caption{The optimal squeezing ratios for nonstationary bistable Gaussian pulses, in the case of $\protect\gamma =-0.1$ and $\protect\alpha =1.3$.
The squeezing ratio for the ordinary nonlinear-Schr\"{o}dinger soliton (the dash-dotted line, $\protect\gamma =0.0$) is plotted for comparison.}
\label{fig-srb}
\end{figure}

When $\gamma < 0$, the VA predicts bistable Gaussian pulses for any energy $E_{0}=\alpha |A|^{2}$ \cite{Angelis94}.
Figure (\ref{fig-srb}) shows the optimal squeezing ratios for a pair of the thus defined bistable nonstationary Gaussian pulses, and they are compared with the curve for the NLSE solitons (marked by ``$\gamma =0$'').
This pair of  the bistable Gaussian pulses have a common width, $\alpha =1.3$, but drastically different energies, $E_{0}=1.30$ and $E_{0}=6.66$, respectively.
Like in the case of the bistable solitons belonging to the pair $B$ in Fig. \ref{fig-bi-soliton}, the pulse with the smaller energy shows fluctuations similar to those of the NLSE soliton, while the pulse with the larger energy has a much poorer squeezing ratio.

\section{Conclusion}
\label{secCCL}
In this work, we have applied the back-propagation method to study the quantum fluctuations around bistable solitons described by the cubic-quintic nonlinear Schr\"{o}dinger equation.
It was found that the squeezing ratio for the soliton strongly depends on the sign of the quintic nonlinearity: the self-focusing quintic term helps to squeeze the fluctuations, while the self-defocusing one makes the squeezing poorer.
In particular, the quantum fluctuations around bistable solitons seem totally different due to the self-defocusing quintic nonlinearity, although the solitons have the same width.
Quantum fluctuation about nonstationary Gaussian pulses were also explored; in that case, radiation loss makes the squeezing weaker.

\section*{Acknowledgment}
B.A.M. appreciates hospitality of the Institute of Electro-Optical Engineering at the National Chiao-Tung University (Hsinchu, Taiwan).


\begin{thebibliography}{99}
\bibitem{Edmundson} D.E. Edmundson and R.H. Enns,
Opt. Lett. \textbf{17}, 586 (1992).

\bibitem{Enns}
R.H. Enns and S.S. Rangnekar,
Phys. Rev. A \textbf{45}, 3354 (1992).

\bibitem{saturable-bistable}
R.H. Enns, D.E. Edmundson, S.S. Rangnekar, and A.E. Kaplan,
Opt. Quant. Electr. \textbf{24}, S1295 (1992).

\bibitem{Kaplan85a} A. E. Kaplan, 
\prl {\bf 55}, 1291 (1985).

\bibitem{Kaplan85b} A. E. Kaplan,
IEEE J. Quant. Electron. \textbf{QE-21}, 1538 (1985).

\bibitem{Herrmann92} J. Herrmann, Opt. Comm. \textbf{87}, 161 (1992).

\bibitem{Desyatnikov00} A. Desyatnikov, A. Maimistov, and B. Malomed,
\pre {\bf 61}, 3107 (2000).

\bibitem{Mihalache03} D. Mihalache, D. Mazilu, I. Towers, B. A. Malomed, and F. Lederer,
\pre {\bf 67}, 056608 (2003).

\bibitem{glass} F. Smektala, C. Quemard, V. Couderc, and A. Barth\'{e}l\'{e}my,
J. Non-Cryst. Solids \textbf{274}, 232 (2000).

\bibitem{organic} C. Zhan, D. Zhang, D. Zhu, D. Wang, Y. Li, D. Li, Z. Lu, L. Zhao, and Y. Nie,
J. Opt. Soc. Am. B \textbf{19}, 369 (2002).

\bibitem{Kolokolov} A. A. Kolokolov, Izv. Vyssh,  Uchebn. Zaved.,
Radiofiz, \textbf{17}, 1332 (1994) (in Russian).

\bibitem{Drummond87} P. D. Drummond and S. J. Carter,
\josab {\bf 4}, 1565 (1987).

\bibitem{Lai89a} Y. Lai and H. A. Haus, \pra {\bf 40}, 844 (1989).

\bibitem{Lai89b} Y. Lai and H. A. Haus, \pra {\bf 40}, 854 (1989).

\bibitem{Schmidt00} E. Schmidt, L. Kn{\"{o}}ll, D.-G. Welsch, M. Zielonka, F. K{\"{o}}nig, and A. Sizmann,
\prl {\bf 85}, 3801 (2000).

\bibitem{Lai90} Y. Lai and H. A. Haus, \pra {\bf 42}, 2925 (1990).

\bibitem{Lee04} R.-K. Lee and Y. Lai, \pra in press.

\bibitem{YLai95} Y. Lai and S.-S. Yu,
\pra {\bf 51}, 817 (1995).

\bibitem{Haus90} H. A. Haus and Y. Lai,
\josab {\bf 7}, 386 (1990).

\bibitem{Matsko} A. B. Matsko and V. V. Kozlov,
\pra {\bf 62}, 033811 (2000).

\bibitem{Lai93} Y. Lai,
\josab {\bf 10}, 475 (1993).

\bibitem{Malomed02} B. A. Malomed, \textit{\textquotedblleft Variational
methods in nonlinear fiber optics and related fields\textquotedblright } in
\textit{Progress in Optics, vol.} \textbf{43}, edited by E. Wolf (Elsevier, Amsterdam, 2002).

\bibitem{Angelis94} C. De Angelis,
IEEE J. Quantum Electron. \textbf{QE-21}, 818 (1994).

\end{thebibliography}
\end{document}